\documentclass[%
reprint,
amsmath,amssymb,
aps,
prc,
]{revtex4-2}

\usepackage[dvipdfmx]{graphicx}% Include figure files
\usepackage{dcolumn}% Align table columns on decimal point
\usepackage{bm}% bold math
\usepackage[mathlines]{lineno}% Enable numbering of text and display math
\modulolinenumbers[5]
\usepackage[version=4]{mhchem}
\usepackage{color}

\begin{document}
	
	%\preprint{APS/123-QED}
	\title{\boldmath Search for particle-stable tetraneutrons in thermal fission of \ce{^235U}}% Force line breaks with \\
	
	\author{Hiroyuki Fujioka}
	\email{fujioka@phys.titech.ac.jp}
	\author{Ryutaro Tomomatsu}
	\affiliation{%
		Department of Physics, Tokyo Institute of Technology, Meguro, Tokyo 152-8551, Japan}%
	
	\author{Koichi Takamiya}
	\affiliation{
		Institute for Integrated Radiation and Nuclear Science, Kyoto University, Kumatori, Osaka 590-0494, Japan}
	
	\date{\today}% It is always \today, today,
	
	\begin{abstract}
		\begin{description}
			\item[Background] The existence of a tetraneutron comprising four neutrons has long been debated.
			\item[Purpose] Motivated by a recent observation of particle-stable tetraneutrons,  we investigated potential particle-stable tetraneutron emission in thermal neutron-induced \ce{^235U} fission using a nuclear research reactor.
			\item[Methods] We performed $\gamma$-ray spectroscopy  for a \ce{^88SrCO_3} sample irradiated in a reactor core.  Stable \ce{^88Sr} was expected to produce \ce{^91Sr} by a tetraneutron-induced $(\ce{^4n},n)$ reaction; hence, observation of $\gamma$-rays followed by $\beta$ decay of \ce{^91Sr} would indicate particle-stable tetraneutron emission.
			\item[Results] The $\gamma$-ray spectrum of an irradiated \ce{^88SrCO_3} sample
			did not show any photopeak for \ce{^91Sr}.	
			\item[Conclusion] The emission rate of particle-stable tetraneutrons, if they exist, is estimated to be lower than $8\times 10^{-7}$ per fission at the 95\% confidence level, assuming the cross-sections of reactions induced by hypothetical particle-stable tetraneutrons.
		\end{description}
	\end{abstract}
	
	\maketitle
	%\tableofcontents
	\section{Introduction}
	Atomic nuclei composed of $Z$ protons and $N$ neutrons
	can be particle-stable systems, depending on $Z$ and $N$.
	For each element with a fixed $Z$,
	a stable isotope satisfies $N\sim Z$; however,
	particle stability against neutron emission has a limit, which is known as the neutron drip line.
	A neutron drip line up to neon ($N=10$) has been established~\cite{PhysRevLett.123.212501}.
	A highly debated question is whether a charge-neutral ($Z=0$) 
	multi-neutron system can exist~\cite{Marques:2021mqf};
	extensive searches for tetraneutrons (\ce{^4n}) have been conducted using various methods.
	
	In 2002, evidence of a bound tetraneutron formed in the breakup of \ce{^14Be} was reported~\cite{Marques:2001wh};
	a neutral particle was detected in a liquid scintillator with a signal significantly larger than that expected in the detection of a neutron.  
	A \ce{^10Be} fragment was simultaneously detected leading the authors to conclude a bound tetraneutron was formed in the breakup of $\ce{^14Be}\to \ce{^10Be}+\ce{^4n}$.
	However, an alternative interpretation that two or more neutrons from a resonant tetraneutron might rescatter in the liquid scintillator 
	and cause such a large signal has also been suggested~\cite{Marques:2005vz}.
	
	Later, a double-charge exchange reaction, $\ce{^4He}(\ce{^8He},\ce{^8Be})$, was used to populate tetraneutron states. A peak at $0.83\pm 0.65\pm  1.25\,\mathrm{MeV}$ above the $4n$ emission threshold was observed in the missing-mass spectrum, indicating the existence of a resonant state~\cite{Kisamori:2016jie}, although the statistical and systematic uncertainties made the result also consistent with a bound tetraneutron.
	
	Recently, two experimental groups independently reported the presence of tetraneutrons
	in the form of bound~\cite{Faestermann:2022meh} and resonant~\cite{Duer:2022ehf} states.
	Faestermann~\textit{et al.}~observed a peak structure in the \ce{^10C} energy spectrum for the $\ce{^7Li}(\ce{^7Li},\ce{^10C})4n$ reaction. 
	This was attributed to the formation of a bound tetraneutron state with an energy of $-0.42\pm 0.16\,\mathrm{MeV}$, relative to the four-neutron threshold, along with the first excited state of \ce{^10C}~\cite{Faestermann:2022meh}.
	The negative energy suggests that the tetraneutron is particle-stable.
	Duer~\textit{et al.} performed missing-mass spectroscopy on the $\ce{^8He}(p,p\ce{^4He)}$ reaction
	and observed a resonance-like structure in the missing mass spectrum~\cite{Duer:2022ehf}. 
	They determined the energy and width of the resonant state as $2.37\pm 0.38\pm 0.44\,\mathrm{MeV}$ and $1.75\pm 0.22\pm 0.30\,\mathrm{MeV}$, respectively.
	
	From a theoretical perspective, a general consensus exists that a tetraneutron will not form a bound state 
	if realistic two-body and three-body nuclear forces are used;
	i.e., the unexpected existence of a bound tetraneutron state has a profound impact 
	on our understanding of the many-body neutron force~\cite{Pieper:2003dc,Lazauskas:2005ig,Hiyama:2016nwn}.
	The possible existence of an observable resonant state with a sufficiently narrow width is also under debate~\cite{Marques:2021mqf}.
	For example, the low-energy structures reported in Ref.~\cite{Duer:2022ehf} can be reproduced by
	considering the dineutron--dineutron correlation in the final state without assuming a tetraneutron resonant state~\cite{Lazauskas:2022mvq}.
	
	In this work, we explore the possible emission of a hypothetically bound tetraneutron resulting from thermal neutron-induced fission of \ce{^235U} in a nuclear research reactor.
	The dominant thermal fission process is binary fission, which
	leads to the emission of two heavy nuclear fragments together with 2.4 neutrons, on average; however, 
	ternary fission involving a light nuclear fragment also occurs with a probability of $0.2\%$ per fission~\cite{Wagemans:2004epd, Koster:1999ke}.
	\ce{^4He} is the most frequently emitted nuclide, accounting for $\sim 90\%$ of ternary fission.
	Furthermore, hydrogen isotopes (tritons, protons, and deuterons, in descending order of emission rate),
	unstable helium isotopes (\ce{^6He}, \ce{^8He}), and nuclides with $Z\ge 3$ are emitted in ternary fission. 
	The emission rates of the two-neutron halo nuclei \ce{^11Li} and \ce{^14Be}, 
	relative to those of \ce{^4He}, were measured as $(9.0\pm 2.2)\times 10^{-8}$ and $(3.4\pm 2.0) \times 10^{-8}$
	~\cite{Koster:1999ke}, that is, on the order of $10^{-11}$ per fission.
	Under the assumption that the hypothetically bound tetraneutron is a ternary particle in uranium fission,
	the objective of this study was to evaluate the emission rate per fission.
	
	An emitted tetraneutron is assumed to induce a secondary reaction with the nucleus in a sample irradiated in the reactor core, 
	assuming that it has a sufficiently long half-life to travel to the sample from the fuel rod inside which it was produced.
	For example, a $(\ce{^4n},n)$ reaction converts the stable isotope $^AZ$ into the isotope $^{A+3}Z$.
	The $(\ce{^4n},n)$ reaction may involve three-neutron transfer and/or one-neutron emission preceded by the formation of a compound nucleus $^{A+4}Z^*$.
	If $^{A+3}Z$ is unstable against $\beta$ decay, 
	a $\gamma$-ray with a specific energy is emitted after $\beta$ decay. 
	The observation of $\gamma$-rays from $^{A+3}Z$ indicates the existence of a bound tetraneutron
	unless a competing process forms $^{A+3}Z$ from a nuclide other than the $^AZ$ in the sample, such as an impurity.
	
	The experimental procedure was the same as that used in the well-established instrumental neutron activation analysis (INAA) in radiochemistry~\cite{soete:1972}.
	In INAA, a trace element in a sample under irradiation is activated by neutron capture, i.e., the $(n,\gamma)$ reaction, predominantly with thermal neutrons.
	The concentration of the trace element of interest can be determined using $\gamma$-ray spectroscopy,
	as the thermal neutron flux at the sample position and the neutron-capture cross section are known.
	In contrast to INAA, we determined the ``tetraneutron flux'' at the sample position
	using the concentration of $^AZ$ in the sample and the assumption of a $(\ce{^4n},n)$ reaction cross section.
	
	To the best of our knowledge, 
	only one study on tetraneutrons has been conducted in a research reactor, which was published in 1963~\cite{SCHIFFER1963292}.
	Two types of reactions, $\ce{^14N}(\ce{^4n},n)\ce{^17N}$ and $\ce{^27Al}(\ce{^4n},p2n)\ce{^28Mg}$, were investigated using samples of 3-amino-1,2,4-triazole (\ce{C_2H_4N_4}) and pure aluminum.
	Neither delayed neutrons from \ce{^17N} nor $\gamma$-rays resulting from the \ce{^28Mg} $\beta$ decay
	were detected above the background level.
	The upper limits of tetraneutron emission per fission were estimated as $2\times 10^{-8}$ and $5\times 10^{-9}$, from the neutron and $\gamma$-ray analyses, respectively.
	
	Several attempts have been made to generate bound tetraneutrons or heavier multineutron systems 
	via the spallation of heavy nuclei, such as uranium, 
	using a high-energy beam (for details, see Section 2.2 in Ref.~\cite{Marques:2021mqf}).
	Detraz provided evidence for multineutron-bound states based on the observation of 
	\ce{^72Zn} $\gamma$-rays in a zinc sample after the spallation of tungsten by a $24\,\mathrm{GeV}$ proton beam~\cite{Detraz:1977wz}.
	However, De Boer~\textit{et al.} argued that high-energy tritons produced during spallation can penetrate a shield and induce the $\ce{^70Zn}(t,p)\ce{^72Zn}$ reaction in zinc samples~\cite{DeBoer:1980cab}.
	Evidence of a bound multineutron system with $N\ge 6$ in
	$\alpha$-induced \ce{^238U} fission has also been reported~\cite{Novatsky2012}, 
	based on the observation of $\gamma$-rays from \ce{^92Sr}, which 
	were expected to be produced in a $\ce{^88Sr}({}^N\mathrm{n},(N-4)n)\ce{^92Sr}$ reaction.
	Further, the cross section of photon-induced $\ce{^209Bi}(\gamma, 4n)\ce{^205Bi}$, slightly below and above the threshold, has been measured by observing $\gamma$-rays from \ce{^205Bi}~\cite{Kotanjyan2023}; 
	non-observation of the signal below the threshold and an estimated cross section above the threshold, much higher than predicted one, indicates the production of a resonant tetraneutron state rather than a bound state.
	
	Motivated by the recent indication of bound tetraneutrons~\cite{Faestermann:2022meh},
	we considered revisiting the tetraneutron search in a research reactor worthwhile.
	Contrary to spallation experiments with accelerators, the typical kinetic energy in thermal neutron-induced fission and possible secondary reactions is, at most, on the order of MeV per nucleon.
	Furthermore, a triton cannot produce $^{A+3}Z$ from $^AZ$.
	Therefore, the aforementioned procedure is free of the drawbacks of the spallation method.
	
	Here, we limit our discussion to tetraneutrons.
	Strictly speaking, the isotope $^{A+3}{Z}$ is converted from the stable isotope $^AZ$
	by not only bound tetraneutrons but also bound multineutron systems with $N>4$, 
	including hexaneutrons \ce{^6n} and octaneutrons \ce{^8n},
	which have been investigated far less than tetraneutrons, both experimentally and theoretically~\cite{Marques:2021mqf}.
	In particular, \textit{ab initio} calculations of such systems are lacking.
	Whether the addition of neutrons to tetraneutrons stabilizes the system is an interesting question.
	Whereas a dedicated experiment in search of such an exotic system requires enormous effort, our method can probe any possible multineutron bound systems using equipment for INAA.
	The number of neutrons $N$ cannot be determined directly even if a positive ${}^{A+3}Z$ signal is observed; however, this problem may be remedied through the search for more neutron-rich isotopes, such as ${}^{A+4}Z$, similar to Ref.~\cite{Novatsky2012}.
	
	\section{Experiment}
	The irradiation experiment was conducted using a hydraulic conveyer at the Kyoto University Research Reactor (KUR) in the Institute for Integrated Radiation and Nuclear Science, Kyoto University, with a thermal power of $5\,\mathrm{MW}$ during irradiation. We selected \ce{^88Sr} as the target $^AZ$ for conversion to $^{A+3}Z$ ($\ce{^88Sr}\to\ce{^91Sr}$) via the $(\ce{^4n},n)$ reaction, with a $Q$-value of $20\,\mathrm{MeV}$ minus the binding energy of the tetraneutron. The main reasons for this selection are as follows:
	\begin{enumerate}
		\item For a given $Z$, the number of neutrons of the radioisotope $^{A+3}Z$, i.e., $N+3=A-Z+3$, should be as large as possible 
		to eliminate the possibility of producing $^{A+3}Z$ from stable isotopes by means other than the $(\ce{^4n}, n)$ reaction.
		\ce{^88Sr} has the largest number of neutrons ($N=50$) among the stable isotopes of strontium. 
		\item The cross section for neutron capture ${}^AZ(n,\gamma){}^{A+1}Z$ should be sufficiently small. 
		For each radioisotope, including $^{A+1}Z$, the maximum allowed quantity to be used per day is stipulated by the institute,
		and the irradiation and cooling times should be optimized accordingly.
		\item \ce{^89Sr} is an almost pure $\beta$-emitter, implying that $\gamma$-rays are hardly emitted after \ce{^89Sr} $\beta$ decay.
		Indeed, only $908.96(3)\,\mathrm{keV}$ $\gamma$-rays are emitted with an intensity of $9.56(5)\times 10^{-5}$~\cite{Singh:2013azp}. This feature is beneficial for reducing the dead time of the germanium detector.
		\item The half-life of $^{A+3}Z$ should be at least comparable to the cooling time of the sample after irradiation.
	\end{enumerate}
	
	\begin{figure*}[t]
		\includegraphics[width=\linewidth]{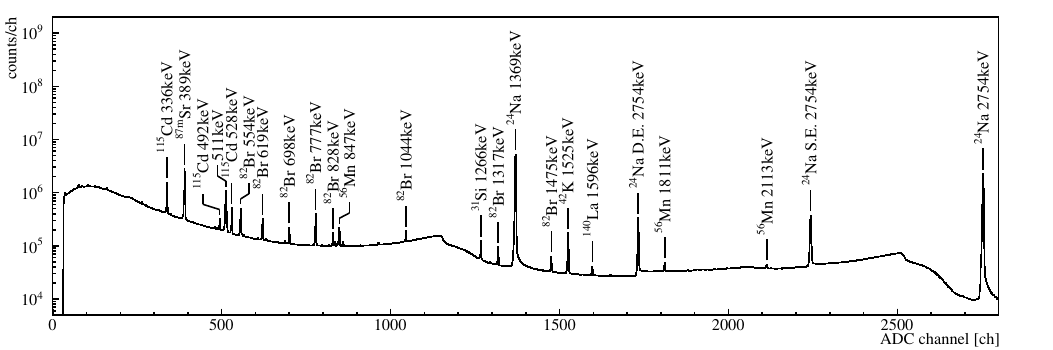}
		\caption{Measured $\gamma$-ray spectrum. The dead-time-corrected spectra of all the measurements over $24\,\mathrm{h}$ were summed. S.E. and D.E. refer to the single and double escape peaks, respectively.}
		\label{spectrum_all}
	\end{figure*}

	As an irradiation sample, $570\,\mathrm{mg}$ of isotopically enriched \ce{^88SrCO_3} (99.90\% enrichment) was enclosed in a silica tube.
	The abundances of the strontium isotopes in the \ce{^88SrCO_3} reagent were less than 0.01\% for \ce{^84Sr}, 0.02\% for \ce{^86Sr}, and 0.08\% for \ce{^87Sr}. 
	The sample was covered with a $0.5\,\mathrm{mm}$-thick cadmium sheet to absorb thermal neutrons and prevent the undesirable activation of the sample owing to a high thermal-neutron capture cross-section of $2\times 10^4\,\mathrm{b}$ for \ce{^113Cd}.
	In contrast, the cadmium sheet would have a negligible impact on a tetraneutron whose kinetic energy is significantly higher than that of thermal neutrons ($\sim 0.025\,\mathrm{eV}$).
	It was then placed in a capsule made of an aluminum alloy for the hydraulic conveyer.
	
	Prior to the experiment, we evaluated two possible contributions to the production of \ce{^91Sr} without tetraneutrons:
	neutron capture of long-lived \ce{^90Sr} and triple neutron capture under the aforementioned conditions.
	For a conservative evaluation, the activation probability was estimated using neutron capture in an irradiation
	without a cadmium absorber, and we ignored the reduction of activated isotopes by $\beta$ decay.
	
	\begin{table}[t]
		\caption{Neutron capture cross sections for thermal neutrons ($\sigma_\text{th}$) and resonance integrals ($I_0$) for \ce{^88Sr}, \ce{^89Sr}, and \ce{^90Sr}.} 
		\label{cross_section}
		\newcolumntype{e}[1]{D{+}{\,\pm\,}{#1}} 
		
		\begin{ruledtabular}
			\renewcommand{\arraystretch}{1.2}
			\begin{tabular}{ccc}
				\textrm{nuclide}&
				$\sigma_\text{th}$ (mb)&
				$I_0$ (mb)\\
				\colrule
				\ce{^88Sr} & $\phantom{00}5.5 \pm \phantom{0}0.4$~\cite{Mughabghab}& $\phantom{0}24\phantom{.0\pm 00}$~\cite{Mughabghab}\\
				\ce{^89Sr} & $420\phantom{.0}\pm 40\phantom{.0}$~\cite{Mughabghab}& $749.8\phantom{{}\pm 00}$~\cite{JENDL}\\
				\ce{^90Sr} & $\phantom{0}10.4\pm \phantom{0}1.4$~\cite{Mughabghab} & $104\phantom{.0}\pm 16$~\cite{Mughabghab}\\
			\end{tabular}
		\end{ruledtabular}
	\end{table}
	
	In general, the activation rate $R$ of neutron capture~\cite{soete:1972} is expressed as
	\begin{align}
		R=\Phi_\text{th}\sigma_\text{th}+\Phi_\text{epi}I_0,\label{eq_activation_rate}
	\end{align}
	where $\Phi_\text{th}$ ($\Phi_\text{epi}$) denotes the thermal (epithermal) neutron flux and $\sigma_\text{th}$ represents the neutron-capture cross section for thermal neutrons. The resonance integral $I_0$ is defined as follows:
	\begin{align}
		I_0=\int_{E_\text{min}}^{E_\text{max}} \frac{\sigma(E)}{E}dE, 
	\end{align}
	where $\sigma(E)$ represents the neutron capture cross section as a function of the neutron energy $E$, and the integration interval $[E_\text{min},E_\text{max}]$ refers to the energy region of epithermal neutrons.
	The nominal neutron flux at the hydraulic conveyer was $\Phi_\text{th}=8.15\times 10^{13}\,(n/\mathrm{cm^2}/\mathrm{sec})$ and 
	$\Phi_\text{epi}=5.95 \times 10^{12}\,(n/\mathrm{cm^2}/\mathrm{sec})$.
	The activation rates during a 2-h irradiation were calculated to be
	$4\times 10^{-9}$ for $\ce{^88Sr}\to \ce{^89Sr}$, $3\times 10^{-7}$ for $\ce{^89Sr}\to \ce{^90Sr}$ and $1\times 10^{-8}$ for $\ce{^90Sr}\to \ce{^91Sr}$
	using the capture cross sections and resonance integrals for strontium isotopes, as summarized in Table~\ref{cross_section}.
	
	A portion of the $\ce{^88SrCO_3}$ reagent was analyzed via energy-filtered thermal ionization mass spectrometry (TIMS)~\cite{TIMS}.
	Consequently, the $\ce{^90Sr}$ signal was not observed, and we concluded that the $\ce{^90Sr}/\ce{^88Sr}$ ratio was below the detection limit of $2.7\times 10^{-12}$~\cite{TIMS, privcom.takagai}.
	As the number of \ce{^88Sr} atoms in the sample was $2.32\times 10^{21}$, the \ce{^91Sr} yield after the 2-h irradiation was estimated to be at most $7\times 10^1$.
	
	Similarly, the probability of triple-neutron capture, i.e., 
	\begin{align}
		\ce{^88Sr}\xrightarrow{(n,\gamma)}\ce{^89Sr}\xrightarrow{(n,\gamma)}\ce{^90Sr}\xrightarrow{(n,\gamma)}\ce{^91Sr},
	\end{align}
	during a 2-h irradiation was estimated to be $2\times 10^{-24}$,
	which corresponds to a \ce{^91Sr} yield of $4\times 10^{-3}$.
	Notably, the triple-neutron capture by \ce{^88Sr} was negligible.
	For example, a similar calculation for the neighboring nuclide \ce{^89Y} resulted in 
	a probability $2\times 10^5$ times higher.
	
	These estimations confirm that the \ce{^88Sr} sample is a promising candidate for investigating bound tetraneutrons via the $(\ce{^4n},n)$ reaction.
	
	The sample was irradiated in a hydraulic conveyer for $2\,\mathrm{h}$, followed by approximately $11\,\mathrm{h}$ of cooling to reduce its radioactivity. 
	We expected that the dominant radioisotopes after cooling would be \ce{^{87\textrm{m}}Sr} (half-life: $2.815(12)\,\mathrm{h}$~\cite{Johnson:2015rrv}) and \ce{^89Sr} (half-life: $50.563 (25)\,\mathrm{d}$~\cite{Singh:2013azp}), with activities of $0.28\,\mathrm{MBq}$ and $0.49\,\mathrm{MBq}$.
	After opening the capsule and removing the highly activated cadmium absorber inside a hot cell, 
	we measured $\gamma$-rays from the activated \ce{^88SrCO_3} sample enclosed in the silica tube
	using a high-purity germanium detector in a hot laboratory.
	The distance between the sample and detector surface was $15\,\mathrm{cm}$.
	We repeated the 30-min measurements 48 times, 
	considering that the half-life of \ce{^91Sr} is $9.65(6)\,\mathrm{h}$~\cite{Baglin:2013qko}.
	
	\section{Results and Discussion}
	Figure~\ref{spectrum_all} shows the $\gamma$-ray spectrum for all the measurements,
	which was obtained by summing the dead-time-corrected spectrum for each measurement.
	In addition to the $389\,\mathrm{keV}$ photopeak of \ce{^{87\textrm{m}}Sr},
	we observed several photopeaks originating from radioisotopes, such as \ce{^82Br}, \ce{^115Cd}, and \ce{^23Na}
	because of the neutron capture by an impurity that was present in the silica or adhered to the surface of the silica tube.
	A calibration source containing \ce{^60Co} and \ce{^137Cs} was used for energy calibration and determination of the absolute photopeak efficiency.
	Furthermore, eight photopeaks of $\ce{^82Br}$ ranging from 554.352(10) to $1474.895(10)\,\mathrm{keV}$~\cite{Tuli:2019vuf} observed in each measurement were used for the \textit{in situ} calibration.
	
	We did not observe photopeaks at $749.8(1)\,\mathrm{keV}$ or at $1024.3(1)\,\mathrm{keV}$
	originating from the \ce{^91Sr} $\beta$ decay; their intensities per \ce{^91Sr} $\beta$ decay are known to be $I_\gamma=23.7(8)\%$ and 33.5(11)\%, respectively~\cite{Baglin:2013qko}.
	However, the sum peak at $1022\,\mathrm{keV}$ due to the annihilation observed in the $\gamma$-ray spectrum
	hindered the statistical estimation of the strength of the neighboring $1024.3(1)\,\mathrm{keV}$ photopeak.
	Therefore, the upper limit of the \ce{^91Sr} yield was determined using the spectrum near $749.8(1)\,\mathrm{keV}$.
	
	\begin{figure}[h]
		\includegraphics[width=\linewidth]{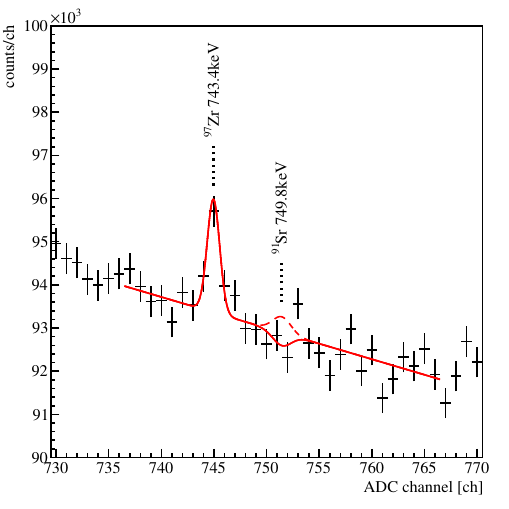}
		\caption{Enlarged $\gamma$-ray spectrum in the vicinity of $749.8\,\mathrm{keV}$. The first 36 measurements were used for the analysis.
			The solid line shows the best fit to the spectrum within a range of $\pm 15\,\mathrm{ch}$ from $751.4\,\mathrm{ch}$ (i.e., $749.8\,\mathrm{keV}$). The negative peak at $751.4\,\mathrm{ch}$ reflects a negative $N_{\ce{^91Sr}}$ in the fitting function (Eq.~(\ref{eq_fitfunc})).  The dashed line indicates the \ce{^91Sr} photopeak with the upper limit of the strength at the 95\% confidence level.}
		\label{spectrum_749}
	\end{figure}
	
	As shown in Fig.~\ref{spectrum_749}, a small \ce{^97Zr} photopeak at $743.36(3)\,\mathrm{keV}$~\cite{Nica:2010niy} was identified near the expected \ce{^91Sr} photopeak.
	Hence, the spectrum was fitted with respect to the two peak functions plus a linear background $aE+b$.
	We found that the shape of the \textit{in situ} $\ce{^82Br}$ photopeaks changed monotonically during each measurement.
	We assumed that the parameters of the shift $\Delta E_j$ and width $\sigma_j$ for the $\ce{^82Br}$ photopeak at $776.511(10)\,\mathrm{keV}$ in the $j^\text{th}$ measurement represented the behavior of the expected \ce{^91Sr} photopeak.
	Moreover, the change in photopeak efficiency during the measurements was corrected for when summing the spectrum from each measurement.
	
	We fitted the summed spectrum from the first measurement until the $N_\text{max}$-th measurement ($1\le N_\text{max} \le 48$) with the following function of the energy $E$ with five free parameters $N_{\ce{^91Sr}}$, $N_{\ce{^97Zr}}$, $\sigma_{\ce{^97Zr}}$, $a$, and $b$:
	\begin{align}
		&f(E; N_{\ce{^91Sr}}, N_{\ce{^97Zr}}, \sigma_{\ce{^97Zr}}, a, b)\nonumber\\ 
		=&\sum_{j=1}^{N_\text{max}} \frac{2^{-(j-1)\Delta t/T_{1/2}}N_{\ce{^91Sr}} }{\sqrt{2\pi}\sigma_j}\exp\left[-\frac{(E-E_{\ce{^91Sr}}-\Delta E_j)^2}{2\sigma_j^2}\right]\nonumber\\
		&\quad{}+\frac{N_{\ce{^97Zr}}}{\sqrt{2\pi}\sigma_{\ce{^97Zr}}}\exp\left[-\frac{(E-E_{\ce{^97Zr}})^2}{2\sigma_{\ce{^97Zr}}^2}\right]+aE+b,\label{eq_fitfunc}
	\end{align}
	where $N_{\ce{^91Sr}}$ represents the area of the \ce{^91Sr} photopeak during the first measurement
	and $N_{\ce{^97Zr}}$ ($\sigma_{\ce{^97Zr}}$) represents the area (width) of the $\ce{^97Zr}$ photopeak 
	in the summed spectrum.
	The literature values of the $\gamma$-ray energies for \ce{^91Sr} and \ce{^97Zr} were input as $E_{\ce{^91Sr}}$ and $E_{\ce{^97Zr}}$, respectively.
	The attenuation of \ce{^91Sr} radioactivity was incorporated into the factor $2^{-(j-1)\Delta t/T_{1/2}}$, 
	where $\Delta t$ represents the measurement time of $0.5\,\mathrm{h}$ and $T_{1/2}$ represents the half-life of \ce{^91Sr}.
	Subsequently, the upper limit of $N_{\ce{^91Sr}}$ at the 95\% confidence level (C.L.) was evaluated
	by integrating the probability density function, which was assumed to be Gaussian in the nonnegative physical region.
	By repeating the fitting for $N_\text{max}=1,\ 2,\ \ldots,\ 48$, 
	the upper limit was minimized to $38.3_{{}-0.7}^{{}+0.8}$ for $N_\text{max}=36$,
	where the systematic error stems from the uncertainty of the photopeak energy, i.e., $\pm 0.1\,\mathrm{keV}$.
	The best fit to the spectrum with $N_\text{max}=36$ and the fit function with $N_{\ce{^91Sr}}$ being the upper limit at the 95\% C.L.
	are displayed in Fig.~\ref{spectrum_749}.
	
	Finally, the upper limit of the \ce{^91Sr} yield in the sample immediately after the 2-h irradiation was determined to be $1.0\times 10^7$ using the intensity of $I_\gamma=23.7(8)\%$ and a photopeak efficiency of $0.116(5)\%$.
	
	To enable comparison with the previous experiment~\cite{SCHIFFER1963292} and
	future experiments at KUR or other research reactors with different configurations,
	we estimated the emission rate of bound tetraneutrons per fission ($R_{\ce{^4n}}$) 
	which serves as an indicator of the experimental sensitivity.
	To relate the \ce{^91Sr} yield after irradiation ($Y_{\ce{^91Sr}}$) to the emission rate,
	we made the same assumptions as in Ref.~\cite{SCHIFFER1963292}:
	\begin{enumerate}
		\item The mean free path of bound tetraneutrons in the light-water moderator was assumed to be $25\,\mathrm{cm}$. Then, the probability that a tetraneutron produced uniformly in the fuel rods strikes a unit area in the sample position was numerically calculated to be $1.0\times 10^{-4}/\mathrm{cm}^2$, using the configuration of the reactor core~\cite{privcom.takahashi}.
		\item The cross section of the $\ce{^88Sr}(\ce{^4n},n)\ce{^91Sr}$ reaction was assumed to be $50\,\mathrm{mb}$, ignoring the difference of the target nucleus. The assumption was made in Ref.~\cite{SCHIFFER1963292} based on the $(\alpha,p)$ and $(\alpha,n)$ cross sections on light nuclei. Whereas this assumption is not grounded in that bound tetraneutrons would be much loosely bound than $\alpha$ particles, and that the Coulomb barrier between a tetraneutron and a target nucleus does not exist, we tentatively adopt the assumption for an order-of-magnitude estimation.
	\end{enumerate}
	Using a fission rate of $1.6\times 10^{17}$ per second, which can be derived by dividing the thermal power of $5\,\mathrm{MW}$ by the released energy of $190\,\mathrm{MeV}$ during fission,
	the ``tetraneutron flux'' at the sample position was estimated to be $1.7\times 10^{13}\times R_{\ce{^4n}}\,\mathrm{/cm^2/s}$.
	By multiplying it by the number of \ce{^88Sr} atoms in the sample and the $(\ce{^4n},n)$ cross section,
	the \ce{^91Sr} yield can be expressed as
	\begin{align}
		Y_{\ce{^91Sr}} = 1.3\times 10^{13}\times R_{\ce{^4n}}.
	\end{align}
	Hence, we determined that the upper limit of $R_{\ce{^4n}}$ was $8\times 10^{-7}$ per fission at the 95\% C.L.
	
	We consider $\gamma$-ray spectroscopy to have an advantage over counting delayed neutrons~\cite{SCHIFFER1963292} because the $\gamma$-ray spectrum contains rich information,
	such as the contributions of impurities.
	Moreover, whereas $\beta$-delayed neutron emitters, e.g., \ce{^17N}, have short lifetimes of $<1$ min,
	our method can be applied to moderately long-lived isotopes such as \ce{^91Sr}, allowing the irradiation time to be significantly increased.
	
	\section{Conclusion}
	We demonstrated the feasibility of particle-stable tetraneutron search using an INAA-like approach in a nuclear reactor.
	In the first measurement with a $570\,\mathrm{mg}$ sample of \ce{^88SrCO_3} irradiated for $2\,\mathrm{h}$ at KUR, $\gamma$ rays from \ce{^91Sr}, which were expected to be produced
	by a tetraneutron-induced $\ce{^88Sr}(\ce{^4n},n)\ce{^91Sr}$ reaction, were not observed.
	Through fitting the $\gamma$-ray spectrum, the yield of \ce{^91Sr} after irradiation
	was deduced to be less than $1.0\times 10^7$ at the 95\% C.L.
	This corresponds to an emission rate of particle-stable tetraneutrons, assuming they exist, of 	less than approximately $8\times 10^{-7}$ per fission at the 95\% C.L., 
	using the $(\ce{^4n},n)$ cross section and mean free path in the moderator conjectured in Ref.~\cite{SCHIFFER1963292}.
	Modern calculations for reactions induced by tetraneutrons or heavier multineutron systems would help quantitative discussions on the emission rate.
	
	In the future, increasing the experimental sensitivity will be crucial.
	For example, extracting irradiated \ce{^88SrCO_3} powder from an activated silica tube
	will reduce the background in $\gamma$-ray spectroscopy.
	Additionally, the number of impurities can be reduced via
	chemical processing of the powder.
	Thus, a subtle signal owing to particle-stable multineutron systems may be observed in high-purity samples irradiated in a nuclear reactor.

	
	\begin{acknowledgments}
		We acknowledge Prof.~Y.~Takagai and Mr.~J.~Aoki at Fukushima University for analyzing the \ce{^90Sr}/\ce{^88Sr} isotope ratio of our sample. 
		We appreciate the staffs of Kyoto University for their cooperation in the irradiation experiment.
		We thank Prof.~Y.~Takahashi at Kyoto University for providing information on the arrangement of fuel rods in the KUR reactor core.
		This work was supported by a Grant for Basic Science Research Projects from the Sumitomo Foundation.
	\end{acknowledgments}
	
	%\nocite{*}
	
	\bibliography{tetraneutron}% Produces the bibliography via BibTeX.
	
\end{document}